\documentclass[aps,preprint]{revtex4}%
\usepackage{amsfonts}
\usepackage{amsmath}
\usepackage{amssymb}
\usepackage{graphicx}%
\providecommand{\U}[1]{\protect\rule{.1in}{.1in}}
\begin{document}
\preprint{ }
\title{Any four orthogonal ququad-ququad maximally entangled states are locally markable}
\author{Li-Yi Hsu}
\affiliation{Department of Physics, Chung Yuan Christian University, Chungli 32081, Taiwan}

\begin{abstract}
In quantum state discrimination, the observers are given a quantum system and
aim to verify its state from the two or more possible target states. In the
local quantum state marking as an extension of quantum state discrimination,
there are N composite quantum systems and $N$ possible orthogonal target
quantum states. Distant Alice and Bob are asked to correctly mark the states
of the given quantum systems via local operations and classical communication.
Here we investigate the local state marking with $N$ $4\otimes4$ systems,
$N=4$, $5$, $6$, and $7$. Therein, Alice and Bob allow for three local
operations: measuring the local observable either $\sigma_{z}$ or $\sigma_{x}$
simultaneously, and entanglement swapping. It shows that, given arbitrary four
$4\otimes4$ systems, Alice and Bob can perform the perfect local quantum state
marking. In the $N=5$, $6$ cases, they can perform perfect local state marking
with specific target states. We conjecture the impossibility of the local
quantum state marking given any seven target states since Alice and Bob cannot
fulfill the task in the simplest case.

\end{abstract}
\volumeyear{year}
\volumenumber{number}
\issuenumber{number}
\eid{identifier}
\startpage{1}
\endpage{2}
\maketitle

\section{Introduction}

In quantum communication, an information carrier is a quantum system with
information encoded in its quantum state, which should be faithfully read out
at the end of the quantum information processing. If classical information is
encoded in non-orthogonal quantum states, the no-cloning theorem prevents the
readability without any ambiguity. Quantum state discrimination (QSD) refers
to the task of distinguishing the state of a given quantum system with a set
of possible target non-orthogonal\ states \cite{qsd,qsdd,qsddd,qsdddd}. There
are two well-known strategies for optimal QSD. One is to minimize the
unavoidable errors of the state distinction \cite{me1,me2,me3,me4}, and the
other is unambiguous discrimination where conclusive result with no error can
be achieved, but an inconclusive answer may occur
\cite{un1,un2,un3,un4,un5,un6}. In the scenario of local state discrimination
(LSD), a multipartite quantum system is distributed among the distant parties
that are tasked with QSD. To distinguish the given quantum state, these
spatially separated parties each are limited to performing the local
operations and classical communication (LOCC). Even when all the possible
states of a composite system are mutually orthogonal, error-free LSD is not
guaranteed. For example, regarding all target states as a special complete set
of orthogonal product states in a two-qutrit system, there is nonlocality
without entanglement that two distant parties cannot implement perfect
distinction by any sequence of LOCC \cite{nwoe}. Further, a set of
multipartite orthogonal quantum states is strongly nonlocal it is locally
irreducible for each bipartition of the subsystems \cite{snwoe}.

During the past two decades, there is considerable effort devoted to a variant
of LSD problems \cite{lsd1,lsd2,lsd3,lsd4,lsd5}. An important result in
\cite{lsd1} states that arbitrary two orthogonal states, entangled or
otherwise, can be locally distinguished with certainty by LOCC. Regarding
quantum entanglement as resource in quantum information processing, local
discrimination of maximally entangled states is studied in great detail. Ghosh
et al. proved three 2-qubit Bell states cannot be distinguished with certainty
by LOCC \cite{lsd2}. In general, \textit{d}+1 or more $d\otimes d$ maximally
entangled states are locally indistinguishable by LOCC \cite{lsd5,lsd6}.
However, $d+1$ is not the tight bound for the number of local
indistinguishability of $d\otimes d$ maximally entangled states. It is
demonstrated that a specific set of four $4\otimes4$ maximally entangled
states are locally indistinguishable by LOCC \cite{qqom}. On the other hand,
if an extra Bell state is supplied as entanglement discrimination catalyst,
perfect LSD on these four orthogonal ququad-ququad maximally entangled states
can be achieved. In \cite{qqom}, a specific set of four $4\otimes4$ maximally
entangled states is investigated. Later a set of six ququad-ququad states is
explored, where any four $4\otimes4$ maximally entangled states in the set are
locally indistinguishable \cite{6}. Very recently, an alternative QSD scheme
termed local quantum state marking (LQSM) is proposed \cite{lqsm}. Unlike LSD
with only a quantum system, in LQSM, there are $N$ states of the given quantum
systems and $N$ possible target states are given. The task of LQSM is to
correctly certify the one-to-one correspondence between the given quantum
systems and the target states with certainty. In the $N=4$ case \cite{lqsm}, a
specific set of four $4\otimes4$ maximally entangled states that are locally
indistinguishable can be correctly marked by a sequence of LOCC.

In this paper, we generalize the result in \cite{lqsm} as follows. \textit{Any
four ququad-ququad maximally entangled states can be locally marked with
certainty.} In the following, four Bell states each can be stabilized by the
operators $\pm\sigma_{x}\otimes\sigma_{x}$ and $\pm\sigma_{z}\otimes\sigma
_{z}$. An unknown Bell state can be distinguished once its eigenvalues of two
joint observables $\sigma_{x}\otimes\sigma_{x}$ and $\sigma_{z}\otimes
\sigma_{z}$ as the two-bit information can be learned simultaneously. A
ququad-ququad maximally entangled state is a product state of two Bell states.
It is important to learn the knowledge of the Bell states only by LOCC while
the quantum measurements result in the consumption of the measured Bell
states. In the strategy of successful LQSM, there are two major ways of
distinguishing two-qubit Bell states. Alice and Bob each usually implement
one-qubit measurement to learn the eigenvalue of the joint observable either
$\sigma_{x}\otimes\sigma_{x}$ or $\sigma_{z}\otimes\sigma_{z}$. In this case,
even though the partial information of an unknown Bell state can be learned,
it can be exploited for the coarse-grained Bell-state classification. The
other way is entanglement swapping. After the coarse-grained classification,
Alice and Bob may infer and know the Bell states of some composite systems
without measurements. Such known and undisturbed Bell states can be exploited
as resource of Bell state discrimination. For example, given a ququad-ququad
state with its half being known, Alice and Bob can definitely verify\ the Bell
state of the other half by entanglement swapping. In this case, Alice and Bob
each perform the Bell state measurement and hence gain full information of the
other unknown Bell state. Given any four\ ququad-ququad maximally entangled
states, there are always sufficient known and undisturbed Bell states consumed
for further fine-grained classification, which is impossible in LSD.
Consequently, Alice and Bob can perform perfect LQSM in $N=4$ case. It will be
shown that only some specific $5$, $6$ ququad-ququad states are locally markable.

\section{Preliminaries and notations}

Denote four 2-qubit Bell states by $\left\vert \overline{xz}\right\rangle $,
$x$, $z\in\{0,1\}$. Explicitly, $\left\vert \overline{00}\right\rangle
=\frac{1}{\sqrt{2}}(\left\vert 00\right\rangle +\left\vert 11\right\rangle )$,
$\left\vert \overline{10}\right\rangle =\frac{1}{\sqrt{2}}(\left\vert
00\right\rangle -\left\vert 11\right\rangle ),\left\vert \overline
{01}\right\rangle =\frac{1}{\sqrt{2}}(\left\vert 01\right\rangle +\left\vert
10\right\rangle )$, and $\left\vert \overline{11}\right\rangle =\frac{1}%
{\sqrt{2}}(\left\vert 01\right\rangle -\left\vert 10\right\rangle )$. Note
that $\left\vert \overline{xz}\right\rangle $\ is a stabilizer state
stabilized by the product operators $(-1)^{x}\sigma_{x}\otimes\sigma_{x}$ and
$(-1)^{z}\sigma_{z}\otimes\sigma_{z}$, where $\sigma_{x}$ and $\sigma_{z}$ are
Pauli matrices. The sets of four ququad-ququad states are defined as
$S_{XZ}=\{\left\vert \overline{x_{1}z_{1}}\right\rangle \otimes\left\vert
\overline{x_{2}z_{2}}\right\rangle |$ $x_{1}+x_{2}=X;z_{1}+z_{2}=Z\}$, where
$X$, $Z\in\{0,1\}$ and hereafter the addition is modulo 2.

In the LQSM scenario with four ququad-ququad maximally entangled states, let
the ququad-ququad state of the $i$-th 4-qubit composite system be%

\begin{equation}
\left\vert \chi^{(i)}\right\rangle =\left\vert \overline{x_{1}^{(i)}%
z_{1}^{(i)}}\right\rangle _{A_{1}^{(i)}B_{1}^{(i)}}\otimes\left\vert
\overline{x_{2}^{(i)}z_{2}^{(i)}}\right\rangle _{A_{2}^{(i)}B_{2}^{(i)}%
}\text{,} \label{x1}%
\end{equation}
where $1\leq i\leq4$, the qubits $A_{1}^{(i)}$ and $A_{2}^{(i)}$ are with
Alice and qubits $B_{1}^{(i)}$ and $B_{2}^{(i)}$ with Bob. In the following,
the qubit indices $A_{1}^{(i)}$, $A_{2}^{(i)}$, $B_{1}^{(i)}$ and $B_{2}%
^{(i)}$ are omitted without causing confusion, and the first and second halves
of $\left\vert \chi^{(i)}\right\rangle $ are denoted by $\left\vert \chi
_{1}^{(i)}\right\rangle =$ $\left\vert \overline{x_{1}^{(i)}z_{1}^{(i)}%
}\right\rangle $ and $\left\vert \chi_{2}^{(i)}\right\rangle =\left\vert
\overline{x_{2}^{(i)}z_{2}^{(i)}}\right\rangle $, respectively. Denote the
target state set $\mathcal{S}^{T}=\{\left\vert \chi_{p}\right\rangle ,$
$\left\vert \chi_{q}\right\rangle ,$ $\left\vert \chi_{r}\right\rangle ,$
$\left\vert \chi_{s}\right\rangle \}$ and the system state set $\mathcal{S}%
^{S}=\{\left\vert \chi^{(1)}\right\rangle ,$ $\left\vert \chi^{(2)}%
\right\rangle ,\left\vert \chi^{(3)}\right\rangle ,$ $\left\vert \chi
^{(4)}\right\rangle \}$. We have (i) $\left\vert \chi^{(i)}\right\rangle
\neq\left\vert \chi^{(j)}\right\rangle \Longleftrightarrow i\neq j$ and (ii)
$\left\vert \chi^{(i)}\right\rangle \in\mathcal{S}^{T}$ $\forall$ $i$,
$j\in\{1,2,3,4\}$. Alice and Bob are tasked to mark the quantum system with
the correct target state with certainty. That is, they are to find the
bijection relation between $\mathcal{S}^{T}$ and $\mathcal{S}^{S}$. Restricted
to LOCC, Alice and Bob allow for three different local quantum operations as follows.

\textit{Entanglement swapping }Given a ququad-ququad state $\left\vert
\overline{x_{1}z_{1}}\right\rangle _{A_{1}B_{1}}\otimes\left\vert
\overline{x_{2}z_{2}}\right\rangle _{A_{2}B_{2}}$ with \textit{ }$\left\vert
\overline{x_{1}z_{1}}\right\rangle =\left\vert \chi_{j}^{(i)}\right\rangle $
and $\left\vert \overline{x_{2}z_{2}}\right\rangle =\left\vert \chi
_{j^{\prime}}^{(i^{\prime})}\right\rangle $, Alice (Bob) performs Bell-state
measurement on qubits $A_{1}$ and $A_{2}$ ($B_{1}$ and $B_{2}$). Note that the
state $\left\vert \overline{x_{1}z_{1}}\right\rangle \otimes\left\vert
\overline{x_{2}z_{2}}\right\rangle $ is a result of combining the $j$ and
$j^{\prime}$ halves from the $i$- and $i^{\prime}$-th composite systems. Let
the post-selected states be $\left\vert \overline{x_{1}^{\prime}z_{1}^{\prime
}}\right\rangle _{A_{1}A_{2}}$ and $\left\vert \overline{x_{2}^{\prime}%
z_{2}^{\prime}}\right\rangle _{B_{1}B_{2}}$, respectively. We have
\begin{equation}
x_{1}+x_{2}=x_{1}^{\prime}+x_{2}^{\prime},\text{ }z_{1}+z_{2}=z_{1}^{\prime
}+z_{2}^{\prime}. \label{x2}%
\end{equation}
For instance, it is easy to verify that\textit{ }$\left\vert \overline
{00}\right\rangle _{A_{1}B_{1}}\otimes\left\vert \overline{00}\right\rangle
_{A_{2}B_{2}}=\frac{1}{2}($\textit{ }$%
%TCIMACRO{\dsum \nolimits_{i\text{, }j=0}^{1}}%
%BeginExpansion
{\displaystyle\sum\nolimits_{i\text{, }j=0}^{1}}
%EndExpansion
\left\vert \overline{ij}\right\rangle _{A_{1}A_{2}}\otimes\left\vert
\overline{ij}\right\rangle _{B_{1}B_{2}})$. As a result, Alice knows the bits
$x_{1}^{\prime}$ and $z_{1}^{\prime}$ and Bob knows the bits $x_{2}^{\prime}$
and $z_{2}^{\prime}$. If the state $\left\vert \overline{x_{1}z_{1}%
}\right\rangle $ is known, Alice and Bob can deduce $x_{2}$, $z_{2}$, and
hence $\left\vert \overline{x_{2}z_{2}}\right\rangle $. To demonstrate the
power of entanglement swapping, let $\left\vert \chi_{p}\right\rangle \in
S_{00}$, $\left\vert \chi_{q}\right\rangle \in S_{01}$, $\left\vert \chi
_{r}\right\rangle \in S_{10}$, $\left\vert \chi_{s}\right\rangle \in S_{11}$.
Alice and Bob perform entanglement swapping on any of these four unknown
ququad-ququad states ($i=i^{\prime}$ and $j=j^{\prime}$). According to the
post-selected states $\left\vert \overline{x_{1}^{\prime}z_{1}^{\prime}%
}\right\rangle _{A_{1}A_{2}}$ and $\left\vert \overline{x_{2}^{\prime}%
z_{2}^{\prime}}\right\rangle _{B_{1}B_{2}}$, they can conclude definitely that
the pre-selected state belongs to the set $S_{(x_{1}^{\prime}+x_{2}^{\prime
})(z_{1}^{\prime}+z_{2}^{\prime})}$ and hence distinct the state with certainty.

\textit{Local Pauli measurements} (LPM) To resolve the eigenvalues of Pauli
product observables, two kinds of local Pauli measurements are exploited.
\textit{(a)} \textit{Local Pauli-X measurements} (LPxM). Alice and Bob each
measure $\sigma_{x}$ on the unknown\textit{ }$\left\vert \overline{x_{i}z_{i}%
}\right\rangle _{A_{i}B_{i}}$ with the local outcomes $o_{x}^{(A)}$ and
$o_{x}^{(B)}\in\{1,-1\}$, respectively. Using two-way communication, Alice and
Bob both learn that $x_{i}=\frac{1-o_{x}^{(A)}o_{x}^{(B)}}{2}$. \textit{(b)}
\textit{Local Pauli-Z measurements }(LPzM) Alice and Bob each measure
$\sigma_{z}$ on the unknown\textit{ }$\left\vert \overline{x_{i}z_{i}%
}\right\rangle _{A_{i}B_{i}}$ with the local outcomes $o_{z}^{(A)}$ and
$o_{z}^{(B)}\in\{1,-1\}$, respectively. Using two-way communication, Alice and
Bob both can $z_{i}=\frac{1-o_{z}^{(A)}o_{z}^{(B)}}{2}$. Note that two Bell
states can be perfectly distinguished using LPM. For example, Alice and Bob
can distinguish two states $\left\vert \overline{0z}\right\rangle $
and\textit{ }$\left\vert \overline{1z^{\prime}}\right\rangle $ ($\left\vert
\overline{x0}\right\rangle $ and \textit{ }$\left\vert \overline{x^{\prime}%
1}\right\rangle $) using LPxM (LPzM).

Some comments are in order. Arbitrary two unknown Bell states can be perfectly
distinguished using LPM that, however, is unable to distinguish three unknown
Bell states \cite{lsd2}. Let the unknown state be $\left\vert \chi
\right\rangle =\left\vert \overline{xz}\right\rangle \otimes\left\vert
\overline{x^{\prime}z^{\prime}}\right\rangle $. Regarding Alice and Bob
performing either LPxM or LPzM on $\left\vert \overline{xz}\right\rangle $,
they can gain one-bit information which is either $x$ or $z$. For further
processing, they can perform either LPxM or LPzM on $\left\vert \overline
{x^{\prime}z^{\prime}}\right\rangle $. Such a two-LPM strategy can be
exploited if these four target states all belong to the same set $S_{XZ}$,
where Alice and Bob can gain the bit values $x$ and $z^{\prime}$ ($x^{\prime}$
and $z$), and then derive the bit values $x^{\prime}=X+x$ and $z=Z+z^{\prime}$
($x=X+x^{\prime}$ and $z^{\prime}=Z+z$).

On the other hand, even though Alice and Bob may or may not assure $\left\vert
\chi\right\rangle $ or $\left\vert \overline{xz}\right\rangle $ after local
operations on $\left\vert \overline{xz}\right\rangle $, it is still possible
to deduce the untouched state $\left\vert \overline{x^{\prime}z^{\prime}%
}\right\rangle $ with certainty. In this case, $\left\vert \overline
{x^{\prime}z^{\prime}}\right\rangle $ can be exploited as resource of Bell
state discrimination. Alice and Bob perform the entanglement swapping on
$\left\vert \overline{x^{\prime}z^{\prime}}\right\rangle $ and an unknown Bell
state $\left\vert \overline{x^{\prime\prime}z^{\prime\prime}}\right\rangle $
to gain full two-bit information of $\left\vert \overline{x^{\prime\prime
}z^{\prime\prime}}\right\rangle $. It will be shown that, if they cannot
certify $\left\vert \overline{xz}\right\rangle $ at first, they can identify
$\left\vert \chi\right\rangle $ by later deduction. It is also possible to
deduce an intact unknown state based on the previous measurement results. In
this case, its undisturbed halves can be exploited as resource of Bell state discrimination.

As the end, denote the Bell-state set by $H_{j}=\{\left\vert \overline
{x_{j}^{(i)}z_{j}^{(i)}}\right\rangle |i=1,2,3,4\}$, $j=1,2$, with the
cardinality denoted by $\left\vert H_{j}\right\vert $. Hereafter the
($\left\vert H_{1}\right\vert $, $\left\vert H_{2}\right\vert $) case
indicates that there are $\left\vert H_{1}\right\vert $ and $\left\vert
H_{2}\right\vert $ elements in the sets $H_{1}$ and $H_{2}$. Without loss of
generality, let $2\leq\left\vert H_{2}\right\vert \leq$ $\left\vert
H_{1}\right\vert \leq4$. For brevity, let the Bell-states $\left\vert
a\right\rangle $, $\left\vert b\right\rangle $, $\left\vert c\right\rangle $,
$\left\vert d\right\rangle \in H_{1}$, and $\left\vert \alpha\right\rangle $,
$\left\vert \beta\right\rangle $, $\left\vert \gamma\right\rangle $,
$\left\vert \delta\right\rangle \in H_{2}$, where the states $\left\vert
a\right\rangle $, $\left\vert b\right\rangle $, $\left\vert c\right\rangle $,
$\left\vert d\right\rangle $ are pairwise different, and similarly for the
states $\left\vert \alpha\right\rangle $, $\left\vert \beta\right\rangle $,
$\left\vert \gamma\right\rangle $, $\left\vert \delta\right\rangle $.

\section{Strategies for perfect marking on four ququad-ququad states}

\textit{(}$2$\textit{, }$2$\textit{) case.} These four states read%

\begin{align}
\left\vert \chi_{p}\right\rangle  &  =\left\vert a\right\rangle \left\vert
\alpha\right\rangle ,\nonumber\\
\left\vert \chi_{q}\right\rangle  &  =\left\vert a\right\rangle \left\vert
\beta\right\rangle ,\nonumber\\
\left\vert \chi_{r}\right\rangle  &  =\left\vert b\right\rangle \left\vert
\alpha\right\rangle ,\nonumber\\
\left\vert \chi_{s}\right\rangle  &  =\left\vert b\right\rangle \left\vert
\beta\right\rangle . \label{x4}%
\end{align}
In this simple case, LPM for distinguishing the states $\left\vert
b\right\rangle $ and $\left\vert a\right\rangle $ ($\left\vert \alpha
\right\rangle $ and $\left\vert \beta\right\rangle $) are performed on the
first (second) half. The non-adaptive two-LPM strategy can be employed. For
example, Alice and Bob can assure that the state of the first half is
$\left\vert a\right\rangle $ and $\left\vert \alpha\right\rangle $ according
to the outcomes of the LPMs on the first and second halves of the state
$\left\vert \chi^{(i)}\right\rangle $ , Alice and Bob mark the state of $i$-th
composite system as $\left\vert \chi_{p}\right\rangle $. That is, $\left\vert
\chi^{(i)}\right\rangle =$ $\left\vert \chi_{p}\right\rangle $.

\textit{(}$4$\textit{, }$1$\textit{) case. }As addressed before, since these
four ququad-ququad states must belong to four different sets $S_{00}$,
$S_{01}$, $S_{10}$, and $S_{11}$, Alice and Bob can mark these unknown states
by performing entanglement swapping. Here the second halves are exploited as
resource of Bell state discrimination.

\textit{(}$4$\textit{, }$4$\textit{) case. }The four unmarked states
\begin{align}
\left\vert \chi_{p}\right\rangle  &  =\left\vert a\right\rangle \left\vert
\alpha\right\rangle ,\nonumber\\
\left\vert \chi_{q}\right\rangle  &  =\left\vert b\right\rangle \left\vert
\beta\right\rangle ,\nonumber\\
\left\vert \chi_{r}\right\rangle  &  =\left\vert c\right\rangle \left\vert
\gamma\right\rangle ,\nonumber\\
\left\vert \chi_{s}\right\rangle  &  =\left\vert d\right\rangle \left\vert
\delta\right\rangle . \label{x5}%
\end{align}
Here the two-LPM adaptive strategies are usually employed, where Alice and Bob
take their local operations based on previous outcomes. Without loss of
generality, let $\left\vert a\right\rangle =\left\vert \overline
{00}\right\rangle $, $\left\vert b\right\rangle =\left\vert \overline
{01}\right\rangle $, $\left\vert c\right\rangle =\left\vert \overline
{10}\right\rangle $, and $\left\vert d\right\rangle =\left\vert \overline
{11}\right\rangle $. Given the $\left\vert \chi^{(i)}\right\rangle $, Alice
and Bob perform LPM on its first half for the coarse-grained classification.
Then they perform conditional LPxM based on the collective local outcomes. If
they learn that $x_{1}^{(i)}=0$ or $1$, they can conclude that $\left\vert
\chi^{(i)}\right\rangle \in\left\{  \left\vert \chi_{p}\right\rangle
,\left\vert \chi_{q}\right\rangle \right\}  $ or $\left\vert \chi
^{(i)}\right\rangle \in\left\{  \left\vert \chi_{r}\right\rangle ,\left\vert
\chi_{s}\right\rangle \right\}  $). As an example of fine-grained
classification, let $\left\vert \chi^{(i)}\right\rangle \in\left\{  \left\vert
\chi_{p}\right\rangle ,\left\vert \chi_{q}\right\rangle \right\}  $ and they
can perform the LPM to distinguish between $\left\vert \alpha\right\rangle
=\left\vert \overline{xz}\right\rangle $ and $\left\vert \beta\right\rangle
=\left\vert \overline{x^{\prime}z^{\prime}}\right\rangle $ as follows. If
$x\neq x^{\prime}$($z\neq z^{\prime}$), they can perform LPxM (LPzM) to learn
$x_{2}^{(i)}$ ($z_{2}^{(i)}$). If the measurement outcomes indicates that
$x_{2}^{(i)}=x$ ($z_{2}^{(i)}=z$), they can verify that $\left\vert \chi
^{(i)}\right\rangle =\left\vert \chi_{p}\right\rangle $. Otherwise,
$\left\vert \chi^{(i)}\right\rangle =$ $\left\vert \chi_{q}\right\rangle .$

As addressed before, if these four ququad-ququad states belong to four
different sets $S_{00}$, $S_{01}$, $S_{10}$, and $S_{11}$, Alice and Bob can
perform entanglement swapping to mark any unknown ququad-ququad state with certainty.

\textit{(}$4$\textit{, }$2$\textit{) case. }There are two subcases. For the
simpler case, let the four unmarked states
\begin{align}
\left\vert \chi_{p}\right\rangle  &  =\left\vert a\right\rangle \left\vert
\alpha\right\rangle ,\nonumber\\
\left\vert \chi_{q}\right\rangle  &  =\left\vert b\right\rangle \left\vert
\alpha\right\rangle ,\nonumber\\
\left\vert \chi_{r}\right\rangle  &  =\left\vert c\right\rangle \left\vert
\beta\right\rangle ,\nonumber\\
\left\vert \chi_{s}\right\rangle  &  =\left\vert d\right\rangle \left\vert
\beta\right\rangle . \label{x6}%
\end{align}
Alice and Bob distinguish between $\left\vert \alpha\right\rangle $ and
$\left\vert \beta\right\rangle $. If the state of the second half is assured
as $\left\vert \alpha\right\rangle $, next they can perform the LPM that
distinguishes between $\left\vert a\right\rangle $ and $\left\vert
b\right\rangle $. In this way, they can correctly mark $\left\vert \chi
_{p}\right\rangle $ and $\left\vert \chi_{q}\right\rangle $ with certainty. On
the other hand, if they assure that the second half is $\left\vert
\beta\right\rangle $, next they perform the LPM that distinguishes between
$\left\vert c\right\rangle $ and $\left\vert d\right\rangle $ with
certainty.\ In this way, they can correctly mark $\left\vert \chi
_{r}\right\rangle $ and $\left\vert \chi_{s}\right\rangle $ with certainty.

In the less trivial ($4$, $2$) case, the four unmarked states are%

\begin{align}
\left\vert \chi_{p}\right\rangle  &  =\left\vert a\right\rangle \left\vert
\alpha\right\rangle ,\nonumber\\
\left\vert \chi_{q}\right\rangle  &  =\left\vert b\right\rangle \left\vert
\alpha\right\rangle ,\nonumber\\
\left\vert \chi_{r}\right\rangle  &  =\left\vert c\right\rangle \left\vert
\alpha\right\rangle ,\nonumber\\
\left\vert \chi_{s}\right\rangle  &  =\left\vert d\right\rangle \left\vert
\beta\right\rangle . \label{x7}%
\end{align}
In the $i$-th operation on $\left\vert \chi^{(i)}\right\rangle $, Alice and
Bob perform LPM on its second half to distinguish between $\left\vert
\alpha\right\rangle $ and $\left\vert \beta\right\rangle $ with the first half
intact. Let Alice and Bob assure $\left\vert \beta\right\rangle $ and hence
they ascertain $\left\vert \chi^{(k)}\right\rangle =\left\vert \chi
_{s}\right\rangle $ in $k$-th operation. Next, $\left\vert d\right\rangle $ is
exploited as discrimination in the following steps.

(i) The $k=1$ case. The second intact halves of the other three states
$\left\vert \chi^{(2)}\right\rangle $, $\left\vert \chi^{(3)}\right\rangle $,
and $\left\vert \chi^{(4)}\right\rangle $ are known as $\left\vert
\alpha\right\rangle $, which can be exploited as resource of Bell state
discrimination. In addition, Alice and Bob then perform entanglement swapping
on the remaining three ququad-ququad states since these three\ ququad-ququad
states each must belong to three different sets among $S_{00}$, $S_{01}$,
$S_{10}$, and $S_{11}$.

(ii) The $k\neq1$ case. Alice and Bob perform entanglement swapping on
$\left\vert d\right\rangle $ and $\left\vert \chi_{1}^{(k-1)}\right\rangle $.
As a result, Alice and Bob can distinguish $\left\vert \chi^{(k-1)}%
\right\rangle $. Finally, Alice and Bob can perform some LPM on the first
halves of the other two states for further distinction.

\textit{(}$3$\textit{, }$3$\textit{) case. }There are two classes. One
includes the following four target states
\begin{align}
\left\vert \chi_{p}\right\rangle =  &  \left\vert a\right\rangle \left\vert
\beta\right\rangle ,\nonumber\\
\left\vert \chi_{q}\right\rangle =  &  \left\vert a\right\rangle \left\vert
\gamma\right\rangle ,\nonumber\\
\left\vert \chi_{r}\right\rangle =  &  \left\vert b\right\rangle \left\vert
\alpha\right\rangle ,\nonumber\\
\left\vert \chi_{s}\right\rangle =  &  \left\vert c\right\rangle \left\vert
\alpha\right\rangle . \label{33}%
\end{align}

In the first class, without loss of generality, let the states $\left\vert
a\right\rangle $\ and $\left\vert b\right\rangle $ have some common eigenvalue
of either $\sigma_{x}\otimes\sigma_{x}$ or $\sigma_{z}\otimes\sigma_{z}$,
different from that of $\left\vert c\right\rangle $. Let $\left\vert
\chi^{(k)}\right\rangle =\left\vert c\right\rangle \left\vert \alpha
\right\rangle =\left\vert \chi_{s}\right\rangle $. The state-marking strategy
can be stated as follows.

Step 1: In the $i$-th operation, $1\leq i\leq k$, Alice and Bob perform LPM to
distinguish $\left\vert c\right\rangle $ from $\left\vert a\right\rangle
$\ and $\left\vert b\right\rangle $ in the first half. For example, let
$\left\vert a\right\rangle =\left\vert \overline{00}\right\rangle $,
$\left\vert b\right\rangle =\left\vert \overline{01}\right\rangle $,
$\left\vert c\right\rangle =\left\vert \overline{11}\right\rangle $. We have
$\sigma_{x}\otimes\sigma_{x}\left\vert a\right\rangle =\left\vert
a\right\rangle $, $\sigma_{x}\otimes\sigma_{x}\left\vert b\right\rangle
=\left\vert b\right\rangle $, and $\sigma_{x}\otimes\sigma_{x}\left\vert
c\right\rangle =-\left\vert c\right\rangle $. In the $i$-th operation, they
perform the LPxM on $\left\vert \chi_{1}^{(i)}\right\rangle $. Alice and Bob
can perfectly mark the state $\left\vert \chi^{(k)}\right\rangle =\left\vert
\chi_{s}\right\rangle $ with its intact second half, $\left\vert
\alpha\right\rangle $ in the $k$-operation. Note that so far the second halves
of the other three unmarked states are all different from each other.

Step 2. If $k\neq4$, in the\ following operation, Alice and Bob perform
entanglement swapping on the states $\left\vert \chi_{2}^{(k)}\right\rangle $
and $\left\vert \chi_{2}^{(k+1)}\right\rangle $. As a result, Alice and Bob
can mark $\left\vert \chi^{(k+1)}\right\rangle $ with certainty. If $k=4$,
Alice and Bob perform entanglement swapping on $\left\vert \chi_{2}%
^{(4)}\right\rangle $ and $\left\vert \chi_{2}^{(3)}\right\rangle $. As a
result, Alice and Bob can correctly mark $\left\vert \chi^{(3)}\right\rangle $
with certainty.

Step 3. Regarding the other two unmarked states, Alice and Bob can correctly
mark them with certainty by performing the LPM on their second halves.

As for the other class of \textit{(}$3$\textit{, }$3$\textit{) case}, it
includes the following four unmarked states.%

\begin{align}
\left\vert \chi_{p}\right\rangle =  &  \left\vert a\right\rangle \left\vert
\alpha\right\rangle ,\nonumber\\
\left\vert \chi_{q}\right\rangle =  &  \left\vert a\right\rangle \left\vert
\beta\right\rangle ,\nonumber\\
\left\vert \chi_{r}\right\rangle =  &  \left\vert b\right\rangle \left\vert
\alpha\right\rangle ,\nonumber\\
\left\vert \chi_{s}\right\rangle =  &  \left\vert c\right\rangle \left\vert
\gamma\right\rangle . \label{x8}%
\end{align}
Denote the set $D=\{\{\left\vert \overline{00}\right\rangle ,\left\vert
\overline{11}\right\rangle \},\{\left\vert \overline{01}\right\rangle
,\left\vert \overline{10}\right\rangle \}\}$. If $\{\left\vert b\right\rangle
,\left\vert c\right\rangle \}\in D$ and $\{\left\vert \beta\right\rangle
,\left\vert \gamma\right\rangle \}\in D$, Alice and Bob can adopt the adaptive
strategy similar to the former case to mark these states with certainty. That
is, since $\left\vert a\right\rangle $\ and $\left\vert c\right\rangle $ must
have some common eigenvalue of either $\sigma_{x}\otimes\sigma_{x}$ or
$\sigma_{z}\otimes\sigma_{z}$, different from that of $\left\vert
b\right\rangle $ and again let $\left\vert \chi^{(k)}\right\rangle =\left\vert
\chi_{r}\right\rangle $. As a result, Alice and Bob go through Step 1 to 3 in
the former case to mark these states with certainty.

On the other hand, let the condition either $\{\left\vert b\right\rangle
,\left\vert c\right\rangle \}\notin D$ or $\{\left\vert \beta\right\rangle
,\left\vert \gamma\right\rangle \}\notin D$ hold. Without loss of generality,
let $\{\left\vert b\right\rangle ,\left\vert c\right\rangle \}\notin D$. In
this case, the states $\left\vert b\right\rangle $\ and $\left\vert
c\right\rangle $ must have some common eigenvalue of either $\sigma_{x}%
\otimes\sigma_{x}$ or $\sigma_{z}\otimes\sigma_{z}$ that is different from
that of $\left\vert a\right\rangle $. Hence Alice and Bob can adopt the
following strategy. In the $j$-operation, Alice and Bob perform LPM on the
first half of $\left\vert \chi^{(j)}\right\rangle $ to distinguish the set the
$\left\vert a\right\rangle $ from states $\left\vert b\right\rangle $\ and
$\left\vert c\right\rangle $. If its first half is $\left\vert a\right\rangle
$ and hence $\left\vert \chi^{(j)}\right\rangle \in\{\left\vert \chi
_{p}\right\rangle $, $\left\vert \chi_{q}\right\rangle \}$, Alice and Bob
perform some LPM on the second half to distinguish between $\left\vert
\alpha\right\rangle $ and $\left\vert \beta\right\rangle $; if the first half
is either $\left\vert b\right\rangle $\ and $\left\vert c\right\rangle $ and
hence $\left\vert \chi^{(j)}\right\rangle \in\{\left\vert \chi_{r}%
\right\rangle $, $\left\vert \chi_{s}\right\rangle \}$, Alice and Bob perform
the some LPM to distinguish between $\left\vert \alpha\right\rangle $ and
$\left\vert \gamma\right\rangle $. As a result, they can correctly mark these
four states with certainty. Finally, we conclude that any four orthogonal
ququad-ququad maximally entangled states are locally markable.

\section{Strategies for perfect marking on five ququad-ququad states}

Unlike the four ququad-ququad state cases, it is not always locally markable
given any five ququad-ququad state. Here we investigate the markability of all
possible five-state cases. In the following discussion, the notation
($\left\vert H_{1}\right\vert $, $\left\vert H_{2}\right\vert _{i_{1}+\cdots
i_{\left\vert H_{2}\right\vert }}$) indicates that the first state (e.g.
$\left\vert \alpha\right\rangle $) appears $i_{1}$ times, the second state
(e.g. $\left\vert \beta\right\rangle $) appears $i_{2}$ times, ..., and the
$\left\vert H_{2}\right\vert $-th state appears$\ i_{\left\vert H_{2}%
\right\vert }$ times. Consequently, $i_{1}+\cdots i_{\left\vert H_{2}%
\right\vert }=5$.

\textit{(}$4$\textit{, }$2_{1+4}$\textit{) case. }These five unknown states
are
\begin{align}
\left\vert \chi_{p}\right\rangle  &  =\left\vert a\right\rangle \left\vert
\alpha\right\rangle ,\nonumber\\
\left\vert \chi_{q}\right\rangle  &  =\left\vert a\right\rangle \left\vert
\beta\right\rangle \nonumber\\
\left\vert \chi_{r}\right\rangle  &  =\left\vert b\right\rangle \left\vert
\beta\right\rangle ,\nonumber\\
\left\vert \chi_{s}\right\rangle  &  =\left\vert c\right\rangle \left\vert
\beta\right\rangle ,\nonumber\\
\left\vert \chi_{t}\right\rangle  &  =\left\vert d\right\rangle \left\vert
\beta\right\rangle \label{x9}%
\end{align}
The strategy is very similar to that taken for the \textit{(}$3$\textit{, }%
$3$\textit{) case }in (\ref{33}). In detail, Alice and Bob perform LPM on to
distinguish between $\left\vert \alpha\right\rangle $ and $\left\vert
\beta\right\rangle $ in the second halves. Let $\left\vert \chi^{(k)}%
\right\rangle =\left\vert a\right\rangle \left\vert \alpha\right\rangle $, and
note the other four states, $\left\vert a\right\rangle \left\vert
\beta\right\rangle $, $\left\vert b\right\rangle \left\vert \beta\right\rangle
$, $\left\vert c\right\rangle \left\vert \beta\right\rangle $, and $\left\vert
d\right\rangle \left\vert \beta\right\rangle $, belong to four different sets
$S_{00}$, $S_{01}$, $S_{10}$, and $S_{11}$. If $k\neq5,$ Alice and Bob take
the following strategy. Therein, $\left\vert \chi_{1}^{(k)}\right\rangle
=\left\vert \alpha\right\rangle $ is exploited as resource of Bell state discrimination.

Step1: In the $j$-operation on the state $\left\vert \chi^{(j)}\right\rangle
$, $j=1,..,k\leq4$, Alice and Bob perform LPMs on its second half that
distinguishes between $\left\vert \alpha\right\rangle $ and $\left\vert
\beta\right\rangle $. In the $k$-operation, Alice can Bob can correctly mark
$\left\vert \chi^{(k)}\right\rangle $ as $\left\vert \chi_{p}\right\rangle $
since they can ascertain that its second half is $\left\vert \beta
\right\rangle $.

Step 2: In the $j$-operation on the state $\left\vert \chi^{(j)}\right\rangle
$ with $j=(k+1),..,5$, Then Alice and Bob correctly mark the $\left\vert
\chi^{(j)}\right\rangle $ by performing entanglement swapping on its first and
second halves.

Step 3: If $k\geq2$, Alice and Bob perform entanglement swapping on
$\left\vert \chi_{1}^{(k)}\right\rangle $ and $\left\vert \chi_{1}%
^{(k-1)}\right\rangle $. They can infer $\left\vert \chi_{1}^{(k-1)}%
\right\rangle $ and then learn $\left\vert \chi^{(k-1)}\right\rangle $ with
certainty. As a result, Alice and Bob can correctly mark the state $\left\vert
\chi^{(k-1)}\right\rangle $. \ 

Step 4: Now Alice and Bob can correctly mark the states $\left\vert
\chi^{(k-1)}\right\rangle $, $\left\vert \chi^{(k)}\right\rangle $, and
$\left\vert \chi^{(k+1)}\right\rangle $ at least, and then there are at most
$l$ ($l=1,2$) intact second halves. Specifically, if $k=3$ and hence $l=1$,
Alice and Bob can correctly deduce the state $\left\vert \chi^{(1)}%
\right\rangle $ since $\left\vert \chi^{(2)}\right\rangle $, $\left\vert
\chi^{(3)}\right\rangle $, $\left\vert \chi^{(4)}\right\rangle $ and
$\left\vert \chi^{(5)}\right\rangle $ can be correctly marked after Steps 3;
if $k=4$ and hence $l=2$, there are some LPM to distinguish between
$\left\vert \chi_{1}^{(1)}\right\rangle $, and $\left\vert \chi_{1}%
^{(2)}\right\rangle $ with certainty, and then $\left\vert \chi^{(1)}%
\right\rangle $, and $\left\vert \chi^{(2)}\right\rangle $ can be correctly marked.

If $k=5$, the above strategy is adapted as follows. In Step 1, Alice and Bob
perform the $j$-operation on $\left\vert \chi_{2}^{(j)}\right\rangle $,
$j=1,..,4$. After these four operations, Alice and Bob can infer that the only
intact state is $\left\vert \chi^{(5)}\right\rangle =\left\vert \chi
_{p}\right\rangle =\left\vert a\right\rangle \left\vert \alpha\right\rangle $.
They exploit the first and second halves of $\left\vert \chi_{p}\right\rangle
$, $\left\vert a\right\rangle $ and $\left\vert \alpha\right\rangle $
respectively, as resource of Bell state discrimination. For example, Alice and
Bob perform entanglement swapping on the states $\left\vert a\right\rangle $
($\left\vert \alpha\right\rangle $) and $\left\vert \chi_{1}^{(4)}%
\right\rangle $ ($\left\vert \chi_{1}^{(3)}\right\rangle $). In this way, they
can mark the states $\left\vert \chi^{(3)}\right\rangle $ and $\left\vert
\chi^{(4)}\right\rangle $ with certainty. Finally, they perform LPM that
distinguishes the first halves of the remaining unmarked states $\left\vert
\chi^{(1)}\right\rangle $ and $\left\vert \chi^{(2)}\right\rangle $.

Before processing further, we introduce the following strategy $\mathfrak{S}$
for perfect state-marking.

Step1: In the $j$-operation on the state $\left\vert \chi^{(j)}\right\rangle
$, $j=1,..,5$, Alice and Bob perform LPM either first or second halves of all
these five states. Based on the measurement outcome and the reordering, these
unmarked states are coarse-grained into two sets: $\mathfrak{C}_{3}%
=\{\left\vert \chi^{(1)}\right\rangle $, $\left\vert \chi^{(2)}\right\rangle
$, $\left\vert \chi^{(3)}\right\rangle \}$ and $\mathfrak{C}_{2}=\{\left\vert
\chi^{(4)}\right\rangle $, $\left\vert \chi^{(5)}\right\rangle \}$.

Step 2: Alice and Bob perform the LPM on the other intact half of $\left\vert
\chi^{(4)}\right\rangle $ that distinguishes between $\left\vert \chi
^{(4)}\right\rangle $ and $\left\vert \chi^{(5)}\right\rangle $. Consequently,
they can correctly mark $\left\vert \chi^{(4)}\right\rangle $ with certainty
and then $\left\vert \chi^{(5)}\right\rangle $ by deduction. In addition, the
intact half of $\left\vert \chi^{(5)}\right\rangle $ is exploited as
discrimination resource in Step 3.

Step 3: Alice and Bob perform entanglement swapping on the intact halves of
$\left\vert \chi^{(5)}\right\rangle $ and $\left\vert \chi^{(3)}\right\rangle
$. Consequently, $\left\vert \chi^{(3)}\right\rangle $ can be correctly marked.

Step 4: Alice and Bob LPM on the intact half of $\left\vert \chi
^{(2)}\right\rangle $ that distinguish between the intact halves of the states
$\left\vert \chi^{(1)}\right\rangle $ and $\left\vert \chi^{(2)}\right\rangle
$. Consequently, they can correctly mark $\left\vert \chi^{(2)}\right\rangle $
and then $\left\vert \chi^{(1)}\right\rangle $ by deduction. It should be
emphasized that, after Step 1, the intact halves in $\mathfrak{C}_{2}$\ and
$\mathfrak{C}_{3}$ must be pairwise different

\textit{(}$4$\textit{, }$2_{2+3}$\textit{) case}. Here the five possible
target states are \
\begin{align}
\left\vert \chi_{p}\right\rangle  &  =\left\vert a\right\rangle \left\vert
\alpha\right\rangle ,\nonumber\\
\left\vert \chi_{q}\right\rangle  &  =\left\vert a\right\rangle \left\vert
\beta\right\rangle \nonumber\\
\left\vert \chi_{r}\right\rangle  &  =\left\vert b\right\rangle \left\vert
\alpha\right\rangle ,\nonumber\\
\left\vert \chi_{s}\right\rangle  &  =\left\vert c\right\rangle \left\vert
\beta\right\rangle ,\nonumber\\
\left\vert \chi_{t}\right\rangle  &  =\left\vert d\right\rangle \left\vert
\beta\right\rangle \label{y1}%
\end{align}
Here Alice and Bob employ strategy $\mathfrak{S}$. In Step 1, Alice and Bob
perform LPMs on all second halves that distinguish between $\left\vert
\alpha\right\rangle $ and $\left\vert \beta\right\rangle $. After the
reordering, the states with the same second half are coarse-grained into the
same state sets. Without loss of generality, after Step 1, these two state
sets are $\mathfrak{C}_{3}=\{\left\vert \chi^{(1)}\right\rangle $, $\left\vert
\chi^{(2)}\right\rangle $, $\left\vert \chi^{(3)}\right\rangle \}=\{\left\vert
\chi_{q}\right\rangle $, $\left\vert \chi_{s}\right\rangle $, $\left\vert
\chi_{t}\right\rangle \}$ and $\mathfrak{C}_{2}=\{\left\vert \chi
^{(4)}\right\rangle $, $\left\vert \chi^{(5)}\right\rangle \}=\{\left\vert
\chi_{p}\right\rangle $, $\left\vert \chi_{r}\right\rangle \}$, respectively.

Next, in Step 2, Alice and Bob Alice and Bob perform LPMs that distinguish
between $\left\vert a\right\rangle $ and $\left\vert b\right\rangle $ on the
first half of $\left\vert \chi^{(4)}\right\rangle $. If Alice and Bob
learn\ $\left\vert \chi_{1}^{(4)}\right\rangle =$ $\left\vert a\right\rangle $
($\left\vert b\right\rangle $), they can ensure that $\left\vert \chi
^{(4)}\right\rangle =\left\vert \chi_{p}\right\rangle $ ($\left\vert \chi
_{r}\right\rangle $) and then deduce that $\left\vert \chi^{(5)}\right\rangle
=\left\vert \chi_{r}\right\rangle $ ($\left\vert \chi_{p}\right\rangle $) with
its first half $\left\vert \chi_{1}^{(5)}\right\rangle =\left\vert
b\right\rangle $ ($\left\vert a\right\rangle $) intact. In Step 3, Alice and
Bob perform entanglement swapping on $\left\vert \chi_{1}^{(5)}\right\rangle $
and the unknown state $\left\vert \chi_{1}^{(3)}\right\rangle $. In this way,
$\left\vert \chi_{1}^{(3)}\right\rangle $and hence $\left\vert \chi
^{(3)}\right\rangle $ can be ensured. Finally, Alice and Bob perform LPM that
distinguishes between $\left\vert \chi_{1}^{(1)}\right\rangle $ and
$\left\vert \chi_{1}^{(2)}\right\rangle $. Finally, Alice and Bob can
correctly mark $\left\vert \chi^{(1)}\right\rangle $ and $\left\vert
\chi^{(2)}\right\rangle $ with certainty.

\textit{(}$4$\textit{, }$4$\textit{) case. }Here the five possible target
states are%

\begin{align}
\left\vert \chi_{p}\right\rangle  &  =\left\vert a\right\rangle \left\vert
\alpha\right\rangle ,\nonumber\\
\left\vert \chi_{q}\right\rangle  &  =\left\vert b\right\rangle \left\vert
\beta\right\rangle ,\nonumber\\
\left\vert \chi_{r}\right\rangle  &  =\left\vert c\right\rangle \left\vert
\gamma\right\rangle ,\nonumber\\
\left\vert \chi_{s}\right\rangle  &  =\left\vert d\right\rangle \left\vert
\delta\right\rangle ,\nonumber\\
\left\vert \chi_{t}\right\rangle  &  =\left\vert a\right\rangle \left\vert
X\right\rangle , \label{y2}%
\end{align}
where $X\in\{\beta,\gamma,\delta\}$\textit{. }Without loss of generality, let
$\left\vert a\right\rangle =\left\vert \overline{00}\right\rangle $,
$\left\vert b\right\rangle =\left\vert \overline{01}\right\rangle \left\vert
c\right\rangle =\left\vert \overline{10}\right\rangle $, and $\left\vert
d\right\rangle =\left\vert \overline{11}\right\rangle $, where $\{\left\vert
a\right\rangle $, $\left\vert b\right\rangle \}\notin D$ and $\{\left\vert
c\right\rangle $, $\left\vert d\right\rangle \}\notin D$. If $X\neq\beta$ and
$X\in\{\gamma,\delta\}$, according to Step 1 of strategy $\mathfrak{S}$, Alice
and Bob perform $\sigma_{x}\otimes\sigma_{x}$ on the first halves to certify
the unknown Bell states belonging to $\{\left\vert a\right\rangle $,
$\left\vert b\right\rangle \}$ or $\{\left\vert c\right\rangle $, $\left\vert
d\right\rangle \}$. After Step 1, these unmarked states are coarse-grained as
two sets $\mathfrak{C}_{3}=\{\left\vert \chi^{(1)}\right\rangle ,\left\vert
\chi^{(2)}\right\rangle $, $\left\vert \chi^{(3)}\right\rangle \}=\{\left\vert
\chi_{p}\right\rangle $, $\left\vert \chi_{q}\right\rangle $, $\left\vert
\chi_{t}\right\rangle \}$ and $\mathfrak{C}_{2}=\{\left\vert \chi
^{(4)}\right\rangle $, $\left\vert \chi^{(5)}\right\rangle \}=\{\left\vert
\chi_{r}\right\rangle $, $\left\vert \chi_{s}\right\rangle \}$; if $X=\beta$,
Alice and Bob perform $\sigma_{z}\otimes\sigma_{z}$ on the second halves to
certify such unknown Bell state belonging to the state set $\{\left\vert
a\right\rangle $, $\left\vert c\right\rangle \}$ or $\{\left\vert
b\right\rangle $, $\left\vert d\right\rangle \}$. After Step 1, these unmarked
states are coarse-grained as two sets $\mathfrak{C}_{3}=\{\left\vert
\chi^{(1)}\right\rangle ,\left\vert \chi^{(2)}\right\rangle $, $\left\vert
\chi^{(3)}\right\rangle \}=\{\left\vert \chi_{p}\right\rangle $, $\left\vert
\chi_{r}\right\rangle $, $\left\vert \chi_{t}\right\rangle \}$ and
$\mathfrak{C}_{2}=\{\left\vert \chi^{(4)}\right\rangle $, $\left\vert
\chi^{(5)}\right\rangle \}=\{\left\vert \chi_{q}\right\rangle $, $\left\vert
\chi_{s}\right\rangle \}$. In Step 2, Alice and Bob perform the LPM that
distinguishes between the untouched halves of $\left\vert \chi^{(4)}%
\right\rangle $ and $\left\vert \chi^{(5)}\right\rangle $ on the untouched
half of $\left\vert \chi^{(4)}\right\rangle $, and hence Alice and Bob can
definitely mark $\left\vert \chi^{(4)}\right\rangle $ and $\left\vert
\chi^{(5)}\right\rangle $. In Step 3, Alice and Bob can correctly mark the
state $\left\vert \chi^{(3)}\right\rangle $ by performing entanglement
swapping on the intact halves of $\left\vert \chi^{(3)}\right\rangle $ and
$\left\vert \chi^{(5)}\right\rangle $. Finally, Alice and Bob can correctly
mark the states in $\mathfrak{C}_{2}$ by performing the LPM that distinguishes
between the intact halves of the states $\left\vert \chi^{(1)}\right\rangle $
and $\left\vert \chi^{(2)}\right\rangle $ on that half of $\left\vert
\chi^{(2)}\right\rangle $.

In the following \textit{(}$k$\textit{, }$3$\textit{)}\ cases with $k=3$, $4$,
there are some unmarkable five-state sets and conditionally markable ones.

\textit{(}$4$\textit{, }$3_{1+1+3}$\textit{) case. } There are two types of
these five target states, where one reads%

\begin{align}
\left\vert \chi_{p}\right\rangle  &  =\left\vert a\right\rangle \left\vert
\alpha\right\rangle ,\nonumber\\
\left\vert \chi_{q}\right\rangle  &  =\left\vert a\right\rangle \left\vert
\beta\right\rangle ,\nonumber\\
\left\vert \chi_{r}\right\rangle  &  =\left\vert b\right\rangle \left\vert
\gamma\right\rangle ,\nonumber\\
\left\vert \chi_{s}\right\rangle  &  =\left\vert c\right\rangle \left\vert
\gamma\right\rangle ,\nonumber\\
\left\vert \chi_{t}\right\rangle  &  =\left\vert d\right\rangle \left\vert
\gamma\right\rangle , \label{1131}%
\end{align}
and the other reads

\bigskip%
\begin{align}
\left\vert \chi_{p}\right\rangle  &  =\left\vert a\right\rangle \left\vert
\alpha\right\rangle ,\nonumber\\
\left\vert \chi_{q}\right\rangle  &  =\left\vert a\right\rangle \left\vert
\gamma\right\rangle ,\nonumber\\
\left\vert \chi_{r}\right\rangle  &  =\left\vert b\right\rangle \left\vert
\beta\right\rangle ,\nonumber\\
\left\vert \chi_{s}\right\rangle  &  =\left\vert c\right\rangle \left\vert
\gamma\right\rangle ,\nonumber\\
\left\vert \chi_{t}\right\rangle  &  =\left\vert d\right\rangle \left\vert
\gamma\right\rangle . \label{1132}%
\end{align}
Without loss of generality, let $\{\left\vert a\right\rangle $, $\left\vert
b\right\rangle \}\notin D$ and hence $\{\left\vert c\right\rangle $,
$\left\vert d\right\rangle \}\notin D$. Similarly, Alice and Bob can LPMs on
the first halves to classify these states in either (\ref{1131}) or
(\ref{1132}) into some three-state set $\mathfrak{C}_{3}$ and the two-state
set $\mathfrak{C}_{2}=\{\left\vert \chi^{(4)}\right\rangle $, $\left\vert
\chi^{(5)}\right\rangle \}=\{\left\vert \chi_{s}\right\rangle $, $\left\vert
\chi_{t}\right\rangle \}$. However, one cannot correctly mark the states
$\left\vert \chi^{(4)}\right\rangle $ and $\left\vert \chi^{(5)}\right\rangle
$ since their intact second halves are the same as $\left\vert \gamma
\right\rangle $. As a result, these five ququad-ququad states in the
\textit{(}$4$\textit{, }$3_{1+1+3}$\textit{) }case are unmarkable.

\textit{(}$4$\textit{, }$3_{1+2+2}$\textit{) case. }Similarly, there are two
types of these five target states, where one reads%

\begin{align}
\left\vert \chi_{p}\right\rangle  &  =\left\vert a\right\rangle \left\vert
\alpha\right\rangle ,\nonumber\\
\left\vert \chi_{q}\right\rangle  &  =\left\vert a\right\rangle \left\vert
\gamma\right\rangle ,\nonumber\\
\left\vert \chi_{r}\right\rangle  &  =\left\vert b\right\rangle \left\vert
\beta\right\rangle ,\nonumber\\
\left\vert \chi_{s}\right\rangle  &  =\left\vert c\right\rangle \left\vert
\alpha\right\rangle ,\nonumber\\
\left\vert \chi_{t}\right\rangle  &  =\left\vert d\right\rangle \left\vert
\beta\right\rangle , \label{431}%
\end{align}
and the other reads%

\begin{align}
\left\vert \chi_{p}\right\rangle  &  =\left\vert a\right\rangle \left\vert
\alpha\right\rangle ,\nonumber\\
\left\vert \chi_{q}\right\rangle  &  =\left\vert a\right\rangle \left\vert
\beta\right\rangle ,\nonumber\\
\left\vert \chi_{r}\right\rangle  &  =\left\vert b\right\rangle \left\vert
\gamma\right\rangle ,\nonumber\\
\left\vert \chi_{s}\right\rangle  &  =\left\vert c\right\rangle \left\vert
\alpha\right\rangle ,\nonumber\\
\left\vert \chi_{t}\right\rangle  &  =\left\vert d\right\rangle \left\vert
\beta\right\rangle . \label{432}%
\end{align}
Notably, at least one of these two conditions $\{\left\vert a\right\rangle $,
$\left\vert b\right\rangle \}\notin D$ and $\{\left\vert a\right\rangle $,
$\left\vert d\right\rangle \}\notin D$ must hold. Let $\{\left\vert
a\right\rangle $, $\left\vert b\right\rangle \}\notin D$ ($\{\left\vert
a\right\rangle $, $\left\vert d\right\rangle \}\notin D$) and hence
$\{\left\vert c\right\rangle $, $\left\vert d\right\rangle \}\notin D$
($\{\left\vert b\right\rangle $, $\left\vert c\right\rangle \}\notin D$ ). In
Step 1 of strategy $\mathfrak{S}$, Alice and Bob can perform LPM on the first
halves to certify the unknown Bell states belonging to the state set
$\{\left\vert a\right\rangle $, $\left\vert b\right\rangle \}$ or
$\{\left\vert c\right\rangle $, $\left\vert d\right\rangle \}$ ($\{\left\vert
a\right\rangle $, $\left\vert d\right\rangle \}$ or $\{\left\vert
b\right\rangle $, $\left\vert c\right\rangle \}$). As a result, five states in
(\ref{431}) can be coarse-grained into two sets $\mathfrak{C}_{3}=\{\left\vert
\chi^{(1)}\right\rangle ,\left\vert \chi^{(2)}\right\rangle ,\left\vert
\chi^{(3)}\right\rangle \}=\{\left\vert \chi_{p}\right\rangle $, $\left\vert
\chi_{q}\right\rangle $, $\left\vert \chi_{r}\right\rangle \}$ ($\{\left\vert
\chi_{p}\right\rangle $, $\left\vert \chi_{q}\right\rangle $,$\left\vert
\chi_{t}\right\rangle \}$) and $\mathfrak{C}_{2}=\{\left\vert \chi
^{(4)}\right\rangle ,\left\vert \chi^{(5)}\right\rangle \}=\{\left\vert
\chi_{s}\right\rangle $, $\left\vert \chi_{t}\right\rangle \}$ ($\{\left\vert
\chi_{r}\right\rangle $, $\left\vert \chi_{s}\right\rangle \}$). In Step 2,
Alice and Bob can correctly mark the states $\left\vert \chi^{(4)}%
\right\rangle $ and $\left\vert \chi^{(5)}\right\rangle $ by performing the
LPM distinguishing between the states $\left\vert \alpha\right\rangle $ and
$\left\vert \beta\right\rangle $ on the second half of $\left\vert \chi
_{2}^{(4)}\right\rangle $. In Step 3, they perform entanglement swapping on
$\left\vert \chi_{2}^{(3)}\right\rangle $ and $\left\vert \chi_{2}%
^{(5)}\right\rangle $, which they can deduce that $\left\vert \chi
^{(3)}\right\rangle $ with certainty. Finally, Alice and Bob can correctly
mark the states $\left\vert \chi^{(1)}\right\rangle $ and $\left\vert
\chi^{(2)}\right\rangle $ after the LPM that distinguish the states on
$\left\vert \chi_{2}^{(1)}\right\rangle $ and on $\left\vert \chi_{2}%
^{(2)}\right\rangle $ is performed on $\left\vert \chi_{2}^{(2)}\right\rangle
$. As a result, these five ququad-ququad states in (\ref{431}) are markable.

On the other hand, the states in (\ref{432}) are conditionally markable.
Specifically, if $\{\left\vert a\right\rangle $, $\left\vert b\right\rangle
\}\notin D$, Alice and Bob can perform state-marking using the strategy
$\mathfrak{S}$ as in the previous \textit{(}$4$\textit{, }$3_{1+1+3}$\textit{)
}case. It is because Alice and Bob can coarse-grain these five ququad-ququad
states\ into the sets $\mathfrak{C}_{3}=\{\left\vert \chi_{p}\right\rangle $,
$\left\vert \chi_{q}\right\rangle $, $\left\vert \chi_{r}\right\rangle \}$ and
$\mathfrak{C}_{2}=\{\left\vert \chi_{s}\right\rangle $, $\left\vert \chi
_{t}\right\rangle \}$ at the end of Step 1. In this case, the states in
(\ref{432}) are markable. However, if $\{\left\vert a\right\rangle $,
$\left\vert b\right\rangle \}\in D$, such coarse-graining is impossible. It is
because we have $\mathfrak{C}_{3}=\{\left\vert \chi_{p}\right\rangle $,
$\left\vert \chi_{q}\right\rangle $, $\left\vert \chi_{s}\right\rangle \}$ or
$\mathfrak{C}_{3}=\{\left\vert \chi_{p}\right\rangle $, $\left\vert \chi
_{q}\right\rangle $, $\left\vert \chi_{t}\right\rangle \}$\bigskip\ after Step
1. However, the second halves of the states in $\mathfrak{C}_{3}$ are not
pairwise different. As a result, the states in (\ref{432}) are unmarkable.

\textit{(}$3$\textit{, }$3$\textit{) case. }There are three different types
for this case:\textit{ (}$3_{1+1+3}$\textit{, }$3_{1+1+3}$\textit{),
(}$3_{1+1+3}$\textit{, }$3_{1+2+2}$), and\textit{ (}$3_{1+2+2}$\textit{,
}$3_{1+2+2}$).

\textit{(}$3_{1+1+3}$\textit{, }$3_{1+1+3}$\textit{)} \textit{case.} Five
states are
\begin{align}
\left\vert \chi_{p}\right\rangle  &  =\left\vert a\right\rangle \left\vert
\alpha\right\rangle ,\nonumber\\
\left\vert \chi_{q}\right\rangle  &  =\left\vert b\right\rangle \left\vert
\alpha\right\rangle ,\nonumber\\
\left\vert \chi_{r}\right\rangle  &  =\left\vert c\right\rangle \left\vert
\alpha\right\rangle ,\nonumber\\
\left\vert \chi_{s}\right\rangle  &  =\left\vert c\right\rangle \left\vert
\beta\right\rangle ,\nonumber\\
\left\vert \chi_{t}\right\rangle  &  =\left\vert c\right\rangle \left\vert
\gamma\right\rangle . \label{311un}%
\end{align}
It is impossible to achieve the perfect marking using strategy $\mathfrak{S}$.
Specifically, it can be easy to verify that either $\mathfrak{C}_{2}=$
$\{\left\vert \chi_{p}\right\rangle ,\left\vert \chi_{q}\right\rangle \}$ or
$\mathfrak{C}_{2}=$ $\{\left\vert \chi_{s}\right\rangle ,\left\vert \chi
_{t}\right\rangle \}$, where the intact halves are the same.

As for the subcase \textit{(}$3_{1+1+3}$\textit{, }$3_{1+2+2}$\textit{)},
these five states are
\begin{align}
\left\vert \chi_{p}\right\rangle  &  =\left\vert a\right\rangle \left\vert
\alpha\right\rangle ,\nonumber\\
\left\vert \chi_{q}\right\rangle  &  =\left\vert a\right\rangle \left\vert
\beta\right\rangle ,\nonumber\\
\left\vert \chi_{r}\right\rangle  &  =\left\vert a\right\rangle \left\vert
\gamma\right\rangle ,\nonumber\\
\left\vert \chi_{s}\right\rangle  &  =\left\vert b\right\rangle \left\vert
\alpha\right\rangle ,\nonumber\\
\left\vert \chi_{t}\right\rangle  &  =\left\vert c\right\rangle \left\vert
\beta\right\rangle . \label{113122Con}%
\end{align}
These states are conditionally markable. In details, if $\{\left\vert
b\right\rangle $, $\left\vert c\right\rangle \}\notin D$, Alice and Bob can
perform the LPM that distinguish the state $\left\vert a\right\rangle $ from
$\left\vert b\right\rangle $ and $\left\vert c\right\rangle $. As a result,
the states in (\ref{113122Con}) after Step 1 of the strategy $\mathfrak{S}$
can be classified into two sets $\mathfrak{C}_{3}=\{\left\vert \chi
_{p}\right\rangle $, $\left\vert \chi_{q}\right\rangle $, $\left\vert \chi
_{r}\right\rangle \}$ and $\mathfrak{C}_{2}=\{\left\vert \chi_{s}\right\rangle
,$ $\left\vert \chi_{t}\right\rangle \}$, which Alice and Bob can achieve
perfect state-marking through Steps 2 to Step 4. On the other hand, if
$\{\left\vert b\right\rangle $, $\left\vert c\right\rangle \}\in D$, Alice and
Bob cannot coarse-grain these states in Step 1 and hence perfect state marking
is impossible.

\textit{(}$3_{1+2+2}$\textit{, }$3_{1+2+2}$\textit{) case. }There are two
types. The five states in one type are%

\begin{align}
\left\vert \chi_{p}\right\rangle  &  =\left\vert a\right\rangle \left\vert
\alpha\right\rangle ,\nonumber\\
\left\vert \chi_{q}\right\rangle  &  =\left\vert a\right\rangle \left\vert
\beta\right\rangle ,\nonumber\\
\left\vert \chi_{r}\right\rangle  &  =\left\vert b\right\rangle \left\vert
\alpha\right\rangle ,\nonumber\\
\left\vert \chi_{s}\right\rangle  &  =\left\vert b\right\rangle \left\vert
\beta\right\rangle ,\nonumber\\
\left\vert \chi_{t}\right\rangle  &  =\left\vert c\right\rangle \left\vert
\gamma\right\rangle . \label{a}%
\end{align}
where the condition either $\{\left\vert a\right\rangle $, $\left\vert
c\right\rangle \}\notin D$ or $\{\left\vert b\right\rangle $, $\left\vert
c\right\rangle \}\notin D$ must hold. Without loss of generality, let
$\{\left\vert b\right\rangle $, $\left\vert c\right\rangle \}\notin D$, Alice
and Bob perform the LPM distinguishing the state $\left\vert a\right\rangle $
from $\left\vert b\right\rangle $ and $\left\vert c\right\rangle \}$ on the
first halves of these ququad-ququad states that distinguish the state that in
Step 1 of strategy $\mathfrak{S}$. Consequently, the states in (\ref{a}) can
be coarse-grained as two sets $\mathfrak{C}_{3}=$ $\{\left\vert \chi
_{r}\right\rangle ,$ $\left\vert \chi_{s}\right\rangle ,$ $\left\vert \chi
_{t}\right\rangle \}$ and $\mathfrak{C}_{2}=\{\left\vert \chi_{p}\right\rangle
,$ $\left\vert \chi_{q}\right\rangle \}$. As a result, Alice and Bob can
achieve perfect state-marking through Steps 2 to Step 4 since the intact
halves in the same set are pairwise different.

On the other hand, these five states in the other type are%

\begin{align}
\left\vert \chi_{p}\right\rangle  &  =\left\vert a\right\rangle \left\vert
\alpha\right\rangle ,\nonumber\\
\left\vert \chi_{q}\right\rangle  &  =\left\vert a\right\rangle \left\vert
\beta\right\rangle ,\nonumber\\
\left\vert \chi_{r}\right\rangle  &  =\left\vert b\right\rangle \left\vert
\alpha\right\rangle ,\nonumber\\
\left\vert \chi_{s}\right\rangle  &  =\left\vert b\right\rangle \left\vert
\gamma\right\rangle ,\nonumber\\
\left\vert \chi_{t}\right\rangle  &  =\left\vert c\right\rangle \left\vert
\beta\right\rangle . \label{b}%
\end{align}

If $\{\left\vert b\right\rangle $, $\left\vert c\right\rangle \}\notin D$,
similarly Alice and Bob can perform the LPM that distinguishes the state
$\left\vert a\right\rangle $ from the states $\left\vert b\right\rangle $ and
$\left\vert c\right\rangle \}$ in Step 1 of strategy $\mathfrak{S}$, where the
states in (\ref{b}) can be coarse-grained as two sets $\mathfrak{C}_{3}=$
$\{\left\vert \chi_{r}\right\rangle ,$ $\left\vert \chi_{s}\right\rangle ,$
$\left\vert \chi_{t}\right\rangle \}$ and $\mathfrak{C}_{2}=$ $\{\left\vert
\chi_{p}\right\rangle ,$ $\left\vert \chi_{q}\right\rangle \}$. However, if
$\{\left\vert b\right\rangle $, $\left\vert c\right\rangle \}\in D$, we have
$\{\left\vert a\right\rangle $, $\left\vert c\right\rangle \}\notin D$. Alice
and Bob can perform the LPM distinguishing the state $\left\vert
b\right\rangle $ from the states $\left\vert a\right\rangle $ and $\left\vert
c\right\rangle \}$ in Step 1 of strategy $\mathfrak{S}$. Although they can
reach the coarse-graining state sets $\mathfrak{C}_{3}=$ $\{\left\vert
\chi_{p}\right\rangle ,$ $\left\vert \chi_{q}\right\rangle ,\ \left\vert
\chi_{t}\right\rangle \}$ and $\mathfrak{C}_{2}=\{\left\vert \chi
_{r}\right\rangle ,$ $\left\vert \chi_{s}\right\rangle \}$. However, two of
the second halves of the states in $\mathfrak{C}_{3}$ are in the same state,
$\left\vert \beta\right\rangle $, and the perfect state-marking is impossible.

\section{Discussion}

We have thoroughly investigated local state-marking of $N$ ququad-ququad
states with $N=4$, $5$. The state-marking of six and more ququad-ququad states
is more complicated and beyond our scope. Here we explore the simplest case
with $\left\vert H_{2}\right\vert =2$ for $N=6$ and 7. In the case of
unmarkable five ququad-ququad states, the remaining intact Bell states in the
set either $\mathfrak{C}_{3}$ or $\mathfrak{C}_{2}$ are the same and fail the
fine-graining from Step 2 to Step 4. It will be shown that even in the
simplest case of 7 ququad-ququad states, perfect state-marking is impossible.
It is because there are three unknown Bell states in the set either
$\mathfrak{C}_{3}$ or $\mathfrak{C}_{4}$, which is impossible to distinguish
them perfectly. On this basis, we conjecture that perfect state-marking of any
7 ququad-ququad states is impossible.

As for $N=6$, given the \textit{(}$k$\textit{, }$2_{3+3}$) case with $k=3,4$,
the strategy $\mathfrak{S}$ can be further exploited for the perfect
state-marking. Therein, these six ququad-ququad states read
\begin{align}
\left\vert \chi_{p}\right\rangle  &  =\left\vert a\right\rangle \left\vert
\alpha\right\rangle ,\nonumber\\
\left\vert \chi_{q}\right\rangle  &  =\left\vert b\right\rangle \left\vert
\alpha\right\rangle ,\nonumber\\
\left\vert \chi_{r}\right\rangle  &  =\left\vert c\right\rangle \left\vert
\alpha\right\rangle ,\nonumber\\
\left\vert \chi_{s}\right\rangle  &  =\left\vert a\right\rangle \left\vert
\beta\right\rangle ,\nonumber\\
\left\vert \chi_{t}\right\rangle  &  =\left\vert b\right\rangle \left\vert
\beta\right\rangle ,\nonumber\\
\left\vert \chi_{u}\right\rangle  &  =\left\vert x\right\rangle \left\vert
\beta\right\rangle , \label{y6}%
\end{align}
where $\left\vert x\right\rangle =\left\vert c\right\rangle $ if $k=3$ and
$\left\vert x\right\rangle =\left\vert d\right\rangle $ if $k=4$. For
state-marking, Alice and Bob perform the LPM on the second halves that
distinguish $\left\vert \alpha\right\rangle $ and $\left\vert \beta
\right\rangle $. Without loss of generality, let Alice and Bob perform LPM on
the second halves of these five states $\left\vert a\right\rangle \left\vert
\alpha\right\rangle ,\left\vert b\right\rangle \left\vert \alpha\right\rangle
,\left\vert c\right\rangle \left\vert \alpha\right\rangle ,\left\vert
a\right\rangle \left\vert \beta\right\rangle ,\left\vert b\right\rangle
\left\vert \beta\right\rangle $. They can deduce that the second half of the
only unmeasured ququad-ququad state is $\left\vert \beta\right\rangle $, and
then perform entanglement swapping on $\left\vert x\right\rangle \left\vert
\beta\right\rangle $ to distinguish the state $\left\vert x\right\rangle $.
They can correctly mark the state $\left\vert x\right\rangle \left\vert
\beta\right\rangle $. In addition, they can coarse-grain these five states
into the sets $\mathfrak{C}_{3}=\left\vert \chi^{(1)}\right\rangle ,\left\vert
\chi^{(2)}\right\rangle ,\left\vert \chi^{(3)}\right\rangle \}=\{\left\vert
\chi_{p}\right\rangle ,$ $\ \left\vert \chi_{q}\right\rangle ,$ $\left\vert
\chi_{r}\right\rangle \}$ and $\mathfrak{C}_{2}=\{\left\vert \chi
^{(4)}\right\rangle ,\left\vert \chi^{(5)}\right\rangle \}=\{\left\vert
\chi_{s}\right\rangle $, $\left\vert \chi_{t}\right\rangle \}$. At last, Alice
and Bob Alice and Bob can achieve perfect state-marking through Steps 2 to
Step 4. It is impossible to perform perfect state-marking in the simplest
\textit{(}$4$\textit{, }$2_{3+4}$) case of seven ququad-ququad states using
the strategy $\mathfrak{S}$. In detail, these seven ququad-ququad states are
\begin{align}
\left\vert \chi_{p}\right\rangle  &  =\left\vert a\right\rangle \left\vert
\alpha\right\rangle ,\nonumber\\
\left\vert \chi_{q}\right\rangle  &  =\left\vert b\right\rangle \left\vert
\alpha\right\rangle ,\nonumber\\
\left\vert \chi_{r}\right\rangle  &  =\left\vert c\right\rangle \left\vert
\alpha\right\rangle ,\nonumber\\
\left\vert \chi_{s}\right\rangle  &  =\left\vert a\right\rangle \left\vert
\beta\right\rangle ,\nonumber\\
\left\vert \chi_{t}\right\rangle  &  =\left\vert b\right\rangle \left\vert
\beta\right\rangle ,\nonumber\\
\left\vert \chi_{u}\right\rangle  &  =\left\vert c\right\rangle \left\vert
\beta\right\rangle ,\nonumber\\
\left\vert \chi_{v}\right\rangle  &  =\left\vert d\right\rangle \left\vert
\beta\right\rangle . \label{z9}%
\end{align}
Similar to the strategy in \textit{(}$k$\textit{, }$2_{3+3}$) case of six
ququad-ququad states, Alice and Bob perform the LPM on the second halves of
these six unmarked states that distinguish the states $\left\vert
\alpha\right\rangle $ and $\left\vert \beta\right\rangle $. They can deduce
the second half of the remaining intact ququad-ququad state and then perform
entanglement swapping on this state to correctly mark this state, which can be
either $\left\vert X\right\rangle \left\vert \alpha\right\rangle $, $X\in\{a$,
$b$, $c\}$ or $\left\vert Y\right\rangle \left\vert \beta\right\rangle $,
$Y\in\{a$, $b$, $c$, $d\}$. Without loss of generality, let the remaining
intact ququad-ququad state be $\left\vert a\right\rangle \left\vert
\alpha\right\rangle $. In this case, they can coarse-grain the other six
states into the sets $\mathfrak{C}_{4}=\{\left\vert \chi^{(1)}\right\rangle $,
$\left\vert \chi^{(2)}\right\rangle $, $\left\vert \chi^{(3)}\right\rangle $,
$\left\vert \chi^{(4)}\right\rangle \}=\{\left\vert \chi_{s}\right\rangle $,
$\left\vert \chi_{t}\right\rangle $, $\left\vert \chi_{u}\right\rangle $,
$\left\vert \chi_{v}\right\rangle \}$ and $\mathfrak{C}_{2}=\{\left\vert
\chi^{(5)}\right\rangle ,\left\vert \chi^{(6)}\right\rangle \}=\{\left\vert
\chi_{q}\right\rangle $, $\left\vert \chi_{r}\right\rangle \}$. Alice and Bob
can perform the LPM that distinguishes between the states $\left\vert
b\right\rangle $ and $\left\vert c\right\rangle $ on $\left\vert \chi
_{1}^{(5)}\right\rangle $, and then correctly mark the $\left\vert \chi
^{(5)}\right\rangle $ and then learn $\left\vert \chi^{(6)}\right\rangle $ by
deduction in Step 2. Although $\left\vert \chi_{1}^{(6)}\right\rangle $ can be
exploited as discrimination resource to correctly mark the state $\left\vert
\chi^{(4)}\right\rangle $ in Step 3, it is impossible to distinguish the
remaining \textit{three} orthogonal Bell states. On the other hand, let the
remaining intact ququad-ququad state be $\left\vert a\right\rangle \left\vert
\beta\right\rangle $, the coarse-grained sets are $\mathfrak{C}_{3}%
=\{\left\vert \chi_{t}\right\rangle ,$ $\left\vert \chi_{u}\right\rangle ,$
$\left\vert \chi_{v}\right\rangle \}$ and $\mathfrak{C}_{3}^{\prime
}=\{\left\vert \chi_{p}\right\rangle ,$ $\left\vert \chi_{q}\right\rangle ,$
$\left\vert \chi_{r}\right\rangle \}$. Alice and Bob cannot distinguish the
remaining \textit{three} orthogonal Bell states in each of the sets
$\mathfrak{C}_{3}$ and $\mathfrak{C}_{3}^{\prime}$.

In conclusion, we investigate the local quantum state marking of $N$
orthogonal ququad-ququad maximally entangled states with $N$ target states. It
is found that any four ququad-ququad entangled states can be perfectly marked,
but such task can be only achieved given specific sets of five or six
ququad-ququad entangled states. It is shown that in the simplest case of seven
ququad-ququad entangled states cannot be correctly marked. There is another
unexplored local quantum state marking problem as follows. There are $N$
orthogonal ququad-ququad maximally entangled states with $T$ target states,
where $N\neq T$. In this case, some inference without local operations may not
be possible, and it is unclear whether there suffices enough resource of Bell
state discrimination in entanglement swapping for gaining full information of
some unknown Bell states. As the end, let $T=N-1$. An interesting and maybe
simple question arises as follows. Is it possible for Alice and Bob to verify
the unknown ququad-ququad maximally entangled state without the given target
state ?

\section{Acknowledgement}

The work was supported by Grant No. MOST 110-2112-M-033 -006.

\end{document}